\documentclass[12pt]{article}

\usepackage{amssymb}
\usepackage{amsmath}
\usepackage{amsfonts}
\usepackage{graphicx}
\usepackage{subfigure}
\usepackage{color}
\usepackage{dcolumn}
\usepackage{hyperref}
\numberwithin{equation}{section}
\newcommand{\be}{\begin{equation}}
\newcommand{\ee}{\end{equation}}
\newcommand{\bea}{\begin{eqnarray}}
\newcommand{\eea}{\end{eqnarray}}
\newcommand{\nn}{\nonumber}

\def\a{\alpha}      
       
  \def\G{\Gamma}  
    
\def\e{\epsilon}        
\def\f{\phi}

\def\l{\lambda} 
\def\m{\mu}

\def\r{\rho}

\newcommand{\eq}[1]{(\ref{#1})}

\begin{document}

\begin{titlepage}
\thispagestyle{empty}


\vspace{2cm}

\begin{center}
\font\titlerm=cmr10 scaled\magstep4 \font\titlei=cmmi10
scaled\magstep4 \font\titleis=cmmi7 scaled\magstep4 {
\Large{\textbf{Thermal Width of Quarkonium from Holography}
\\}}
\vspace{1.5cm} \noindent{{
Kazem Bitaghsir
Fadafan$^{}$\footnote{e-mail:bitaghsir@shahroodut.ac.ir }, Seyed
Kamal Tabatabaei$^{}$\footnote{e-mail:k.tabatabaei67@yahoo.com }
}}\\
\vspace{0.8cm}

{\it ${}$Physics Department, Shahrood University, Shahrood, Iran\\}

\vspace*{.4cm}

\end{center}

\vskip 2em

\begin{abstract}
From the AdS/CFT correspondence, the effects of charge and finite 't
Hooft coupling correction on the thermal width of a heavy quarkonium
are investigated. To study the charge effect, we consider Maxwell
charge which is interpreted as quark medium. In the case of finite
't Hooft coupling corrections, $\mathcal{R}^4$ terms and
Gauss-Bonnet gravity have been considered, respectively. It is shown
that these corrections affect the thermal width. It is also argued
that by decreasing the 't Hooft coupling, the thermal width becomes
effectively smaller. Interestingly, this is similar to analogous
calculations in weakly coupled plasma.

\end{abstract}

\end{titlepage}

\tableofcontents
\section{Introduction}
The experiments of Relativistic Heavy Ion Collisions (RHIC) have
produced a strongly coupled quark$-$gluon plasma (QGP)(see review
\cite{CasalderreySolana:2011us}). At a qualitative level, the data
indicate that the QGP produced at the LHC is comparably strongly
coupled and it is expected to be better approximated as conformal
than is the case at RHIC \cite{ALICE}. There are no known
quantitative methods to study strong coupling phenomena in QCD which
are not visible in perturbation theory (except by lattice
simulations). A new method for studying different aspects of QGP is
the $AdS/CFT$ correspondence
\cite{CasalderreySolana:2011us,Maldacena:1997re,Gubser:1998bc,Witten:1998qj,
Witten:1998zw}. This method has yielded many important insights into
the dynamics of strongly coupled gauge theories. It has been used to
study hydrodynamical transport quantities at equilibrium and real
time process in non-equilibrium \cite{DeWolfe:2013cua}. Methods
based on $AdS/CFT$ relate gravity in $AdS_5$ space to the conformal
field theory on the four-dimensional boundary \cite{Witten:1998qj}.
It was shown that an $AdS$ space with a black brane is dual to a
conformal field theory at finite temperature \cite{Witten:1998zw}.

In heavy ion collisions at the LHC, heavy quark related observables
are becoming increasingly important \cite{Laine:2011xr}. In these
collisions, one of the main experimental signatures of QGP formation
is melting of quarkonium systems, like $J/\psi$ and excited states,
in the medium \cite{Matsui:1986dk}. They are also useful probes for
QCD phenomenology \cite{Brambilla:2004wf}. The thermal width of
these systems is an important subject in QGP \cite{First}. This
quantity emerges from the imaginary part of the heavy quark
potential, which is related to quarkonium decay processes in the QGP
\cite{Beraudo:2007ky}. In the effective field theory framework,
thermal decay widths have been studied in
\cite{Brambilla:2008cx,Brambilla:2010vq}. It was shown that at
leading order, two different mechanisms contribute to the decay
width, namely Landau damping and singlet-to-octet thermal breakup.
As long as the Debye screening mass is larger than the binding
energy, the former mechanism dominates over the latter. Also from
the elementary process point of view, the Landau-damping mechanism
corresponds to dissociation by inelastic parton scattering and the
singlet-to-octet thermal breakup corresponds to gluon dissociation
\cite{Brambilla:2011sg}. Beyond leading order, these two mechanisms
would be the same.

The analytic estimate of the imaginary part of the binding energy
and the resultant decay width were studied in \cite{Dumitru:2010id}.
The peak position and its width in the spectral function of heavy
quarkonium can be translated into the real and imaginary part of the
potential \cite{Rothkopf:2011db}. Using the soft-wall AdS/QCD model
the finite-temperature effects on the spectral function in the
vector channel have been studied and a similar behavior to lattice
QCD results was found for the in-medium mass shift and the width
broadening of the vector meson \cite{Fujita:2009wc}.

The effect of the imaginary part of the potential on the thermal
widths of the states in both isotropic and anisotropic plasmas has
been studied in \cite{imaginary2,Margotta:2011ta}. This study has
been done by considering the modifications to the Coulombic wave
function of the imaginary part of the potential for $J/\psi,
\Upsilon$ and $\chi_b$ for both an isotropic and anisotropic QGP.
The expectation value of the imaginary part of the quarkonium
potential gives the thermal width, which was obtained analytically
in the case of an isotropic plasma in \cite{Dumitru:2010id}. For the
case of $\Upsilon$ it is on the order of 20-100 MeV, which is
comparable to the decay width of
$\Upsilon \rightarrow e^+ e^-$.%

The thermal width of heavy quarkonium in a hot strongly coupled
isotropic plasma was, from a holographic point of view, initially
studied in \cite{Noronha:2009da}. In this approach, the thermal
width of heavy quarkonium states originates from the effect of
thermal fluctuations due to the interactions between the heavy
quarks and the strongly coupled medium. This is described
holographically by integrating out thermal long wavelength
fluctuations in the path integral of the Nambu-Goto action in a
curved background spacetime. This study was extended to the case of
anisotropic plasma in \cite{Dimitris1,Fadafan:2013bva} and imaginary
potential formula in a general curved background was obtained. It
was also shown that the thermal width is decreased in the presence
of anisotropy and a larger decrease happens along the transverse
plane. This method was revisited in \cite{Finazzo:2013rqy} and
general conditions for the existence of an imaginary part for the
heavy quark potential were obtained. In the context of AdS/CFT,
there are other approaches which can lead to a complex static
potential \cite{Albacete:2008dz,Hayata:2012rw}. In
\cite{Albacete:2008dz}, the extended range of the radial distance
$u$ was studied in such a way that the string world-sheet solutions
of the Wilson loop corresponding to the static, potential become
complex, and therefore the corresponding static potential develops
an imaginary part. The method in \cite{Hayata:2012rw} is based on
the spectral decomposition of the Euclidean Wilson loop and its
analytic continuation to the real time. The imaginary potential in
this method grows linearly with temperature, which is qualitatively
consistent with that obtained in
\cite{Noronha:2009da,Albacete:2008dz}.

In this paper, we study different effects on the thermal width by
considering the effects of the charge and the finite 't Hooft
coupling correction on the hot plasma. To study the charge effect,
we consider a Maxwell charge which can be interpreted as a quark
medium. Therefore on the gravity side, we consider the non-extremal
Reissner-Nordstrom AdS $(RNAdS)$ black hole. The finite 't Hooft
coupling corrections also correspond to $\mathcal{R}^4$ corrections
and Gauss-Bonnet terms, respectively. An understanding of how the
imaginary part of the potential and thermal width of heavy
quarkonium are affected by these corrections may be essential for
theoretical predictions.

Melting of a heavy meson like $J/ \psi$ and excited states like
$\chi_c$ and $\psi'$ in the quark medium have been investigated in
\cite{Fadafan:2012qy}. It was shown that the excited states melt at
higher temperatures. Heavy quarks in the presence of higher
derivative corrections have been studied in \cite{Noronha:2009ia,
Fadafan:2011gm}. Now we continue with considering these effects on
the thermal width of quarkonium.

This paper is organized as follows. In the next section, we will
present an example for the connection between the imaginary part of
the potential and confinement. This example confirms that the
imaginary potential is zero in the confinement phase. We give the
general expressions to study the thermal width in section 3. Also in
this section, we use the general formulas and investigate the
thermal width behavior at finite coupling
and in the presence of a dense medium. In the last section we summarize our results. %

\section{Imaginary potential and confinement}

As was argued in \cite{Finazzo:2013rqy}, the presence of a black
brane is necessary in order to have a non-zero imaginary potential.
In other words, in the absence of a black brane the imaginary part
of the potential vanishes. In this section, we examine this idea in
a theory which exhibits a confinement-deconfinement transition at
some finite temperature $T_c$. We are not going to carry out the
details of the calculations which prove that the imaginary part of
the potential is zero in the confined phase.

One may consider the confining SU(N) gauge theory based on N D4
branes on a circle. In particular, fundamental parameters are an
energy scale in addition to temperature. The vacuum of the theory at
zero temperature confines the color charge. One expects that the
imaginary potential in the low T phase should vanish.

The model that we consider is the model of Witten
\cite{Witten:1998zw} which is based on N D4 branes wrapped on a
circle. In this theory the field theory is a non-supersymmetric
SU(N) Yang-Mills theory that confines at low temperatures
\cite{Witten:1998zw}. Although the theory is different from pure
Yang-Mills theory, it exhibits linear confinement of the quarks in
vacuum. The gravitational background dual to the vacuum is known
analytically to be %
\bea
ds^2&=&(\frac{u}{l})^{3/2}\left(-dt^2+d\r^2+\r^2d\f^2+dx_3^2+f(u)dx_4^2\right)+(\frac{l}{u})^{3/2}
\left(\frac{du^2}{f(u)}+u^2d\Omega_4^2\right) \nn\\
&&F_{(4)}=\frac{2\pi N}{V_4}\e_4,\,\,\,\,\,e^{\f}=g_s(\frac{u}{l})^{3/4},\,\,\,\,l^3=\pi g_s N_c l_s^3.\label{LowT}\eea%
A typical length scale associated to the D4 brane geometry is $l$
which all dimensionful quantities measure in units of $l$. The
volume of the unit $S^4$ and the associated volume form are $V_4$
and $\e_4$, respectively. Also%
\be f(u)=1-(\frac{u_k}{u})^3,\,\,\,\,\, \frac{u_k}{l}=\frac{4l^2}{9R^2}.\ee%
The latter equation follows from demanding the absence of a conical
singularity at the tip of the cigar $u_k$ that is spanned by $x_4$
and $u$. At the high temperature phase, the black hole geometry is
given
by%
\bea
ds^2&=&(\frac{u}{l})^{3/2}\left(-F(u)dt^2+d\r^2+\r^2d\f^2+dx_3^2+dx_4^2\right)+(\frac{l}{u})^{3/2}
\left(\frac{du^2}{F(u)}+u^2d\Omega_4^2\right). \label{HighT}\eea%
Here the blackness function $F(u)$ and the temperature are%
\be F(u)=1-(\frac{u_h}{u})^3,\,\,\,\,\,\frac{u_h}{l}=\frac{16\pi^2 l^2}{9}T^2. \ee%
There is a critical temperature $T_c=\frac{1}{2\pi R}$ where the
theory is confined for $T<T_c$ and the geometry is described by
\eq{LowT}, whereas, for $T>T_c$ the theory is deconfined and is
given by \eq{HighT}. The ratio of $\frac{u_h}{u_k}$ is given by%
\be \frac{u_h}{u_k}=\left(\frac{T}{T_c}\right)^2.\label{uhuk}\ee%
Based on the results of \cite{Finazzo:2013rqy}, If $u_k > u_h$ then
the U-shaped string cannot go past $u_k$ and one cannot consider
fluctuations beyond $u_k$. From AdS/CFT, one may interpret this
condition from \eq{uhuk} in terms of the temperature. Then, more
explicitly, for $T<T_c$ the imaginary potential vanishes, while for
$T>T_c$ it is not zero. As a result, in this geometry, which shows
the confinement-deconfinement transition at $T_c$, the imaginary
potential is zero in the confinement phase.
\section{Thermal width from holography}
In order to compute thermal widths from holography, one should
consider a heavy quark anti-quark pair in the boundary of a black
hole geometry. In this section, we give the general formulas for
calculating the distance between the quark and the antiquark, $L$,
the real part of the potential $Re V_{Q\bar{Q}}$ and the imaginary
part of the potential $\text{Im} V_{Q\bar{Q}}$ in terms of the
geometry coordinates.

We consider the general gravity as follows:%
\be
 ds^2=G_{tt}dt^2+G_{xx}dx_i^2+G_{uu}du^2,\label{general-background}
\ee%
here the metric elements are functions of the radial distance $u$
and $x_i=x,y,z$ are the boundary coordinates. In these coordinates,
the boundary is located at infinity. We study a static quark
antiquark system at the boundary as an open string in the bulk space
from the gauge string duality point of view \cite{Erdmenger:2007cm}.
Using the usual orthogonal Wilson loop which corresponds to the
heavy Q\={Q} pair, and assuming the system to be aligned in the $x$
direction %
\be t=\tau,\quad x=\sigma,\quad
u=u(\sigma),\label{static-gauge} \ee%
one finds the following generic formulas for the heavy meson. In
these formulas $u_*\equiv u(x=0)$ is the deepest point of the
U-shaped string. The metric functions appear as $V(u) \equiv
-G_{tt}G_{xx}$ and $W(u)\equiv -G_{tt}G_{uu}$.

\begin{itemize}
\item{The distance between quark and
antiquark, $L$ is given by \be\label{staticL1}
L=2\,\int_{u_*}^{\infty}du~\left[\frac{V(u)}{W(u)}\left(\frac{V(u)}{V(u_*)}-1\right)\right]
^{-\frac{1}{2}}~~,  \ee }

\item{The real part of heavy quark potential,
$Re V_{Q\bar{Q}}$ is as follows: \bea Re V_{Q\bar{Q}} = \frac{1}{\pi
\a'}\left[\int_{u_*}^{\infty}du~\left(
\left(\frac{1}{W(u)}-\frac{V(u_*)}{V(u)W(u)}\right)^{-1/2}-\sqrt{W_0(u)}\right)-\int_{0}^{u_*}du~\sqrt{W_0(u)}
\right],\nonumber\\\label{staticV} \eea   here
$W_0(u)=W(u\rightarrow \infty)$. }

\item{The imaginary part of the potential is negative and it is given by
\be \text{Im} V_{Q\bar{Q}}=
-\frac{1}{2\sqrt{2}\alpha'}\left[\frac{V'(u_*)}{2V''(u_*)}-\frac{V(u_*)}{V'(u_*)}
\right]~\sqrt{W(u_*)}~.  \label{ImV} \ee%
The derivatives are with respect to $u$. For the special case of $
W(u) = 1$, this formula reduces to the case of an isotropic plasma
\cite{Noronha:2009da}, while $ W(u)\neq 1$ corresponds to the
anisotropic plasma \cite{Dimitris1,Fadafan:2013bva} .}

\end{itemize}

These generic formulas give the related information of the heavy
quarkonium in terms of the metric elements of a background
\eq{general-background}.

To find the $\text{Im}V$, one should express it in terms of the
length $L$ of the Wilson loop instead of $u_*$ using the equation
\eq{staticL1}. There is an important point for long U-shaped strings
because it would be possible to add new configurations
\cite{Bak:2007fk}. Here we are interested mostly in distances $L
T<1$ and do not consider such configurations.

We use a first-order non-relativistic expansion,%
 \be
\Gamma=-<\psi|\text{ImV}_{Q\bar{Q}}|\psi>,\label{gamma} \ee %
to calculate the thermal width of a heavy quarkonium like $\Upsilon$
meson. The imaginary potential is given by \eq{ImV}; also $|\psi>$
is the Coulombic wave function of the Coulomb potential of the heavy
quarkonium. From the holographic point of view, the potential
between infinitely massive quarks in a pure AdS background is a
Coulomb-like potential,%
\be
V_{Q\bar{Q}}(L)=-\frac{4\,\pi^2}{\Gamma(1/4)^4}\,\frac{\sqrt{\lambda}}{L}.\label{ColV}
\ee%
The original calculation of this result comes from considering the
rectangular Wilson loop in the vacuum of strongly coupled N = 4 SYM
theory \cite{Maldacena:1998im}. The generalization to finite
temperature has been done in \cite{Rey}. The effect of higher
derivative corrections on the real part of the potential within the
gauge/gravity duality has been done in \cite{Noronha:2009ia,
Fadafan:2011gm} and it was found that the dissociation length
becomes shorter with the increase of the coupling parameters of the
higher curvature terms. The rotating heavy quarkonium also is
studied in \cite{ Peeters,Antipin,AliAkbari:2009pf,Fadafan:2012qy}.

Regarding the Coulombic potential in \eq{ColV}, applying a potential
model may therefore provide qualitatively useful insight. We
consider the ground-state wave function of $|\psi>$ in a
Coulomb-like potential $V=-A/L$ where $A$ in the case of N=4 SYM
comes from \eq{ColV}. In the ground state of the energy levels of a
bound state of heavy quarks with mass of $m_Q$, the Bohr radius is
defined as $a_0=2/m_Q A$. The wave function
is%
\be |\psi>=\frac{1}{\sqrt{\pi}a_0^{3/2}} \,e^{-L/a_0}. \label{psi}\ee%
By considering different effects on the plasma, the real part of the
potential is not given by just the Coulombic term. But this term
provides the leading contribution for the potential of heavy
quarkonia; therefore, it justifies the use of a Coulomb-like wave
function $|\psi>$ in \eqref{psi} to determine the width. Considering
a finite temperature plasma will not change the coefficient $A$ in
the Coulombic potential. However, by studying the charge and finite
coupling corrections one should modify $A$. We find the potential
using the numerical fitting in this case. More details will be found
in the next sections where we show the behavior of $\Gamma$ in terms
of parameters of the plasma. As a result, by considering different
effects on the plasma, we also find $A$ by fitting methods. This
approach is carefully followed in \cite{Fadafan:2013bva}.

\begin{figure}
\centerline{\includegraphics[width=3in]{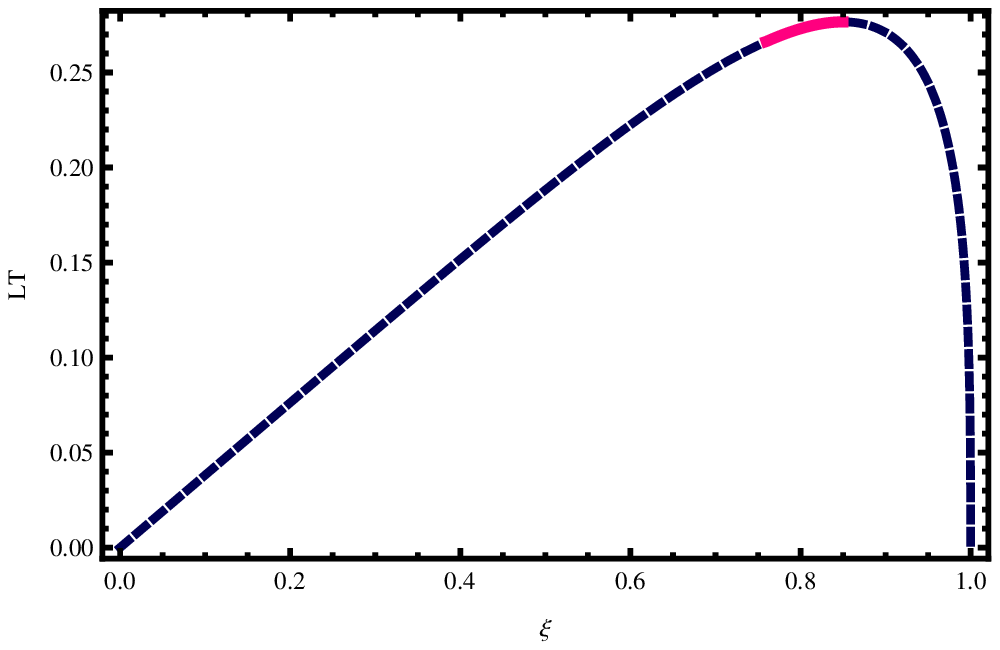}\,\,\includegraphics[width=3in]{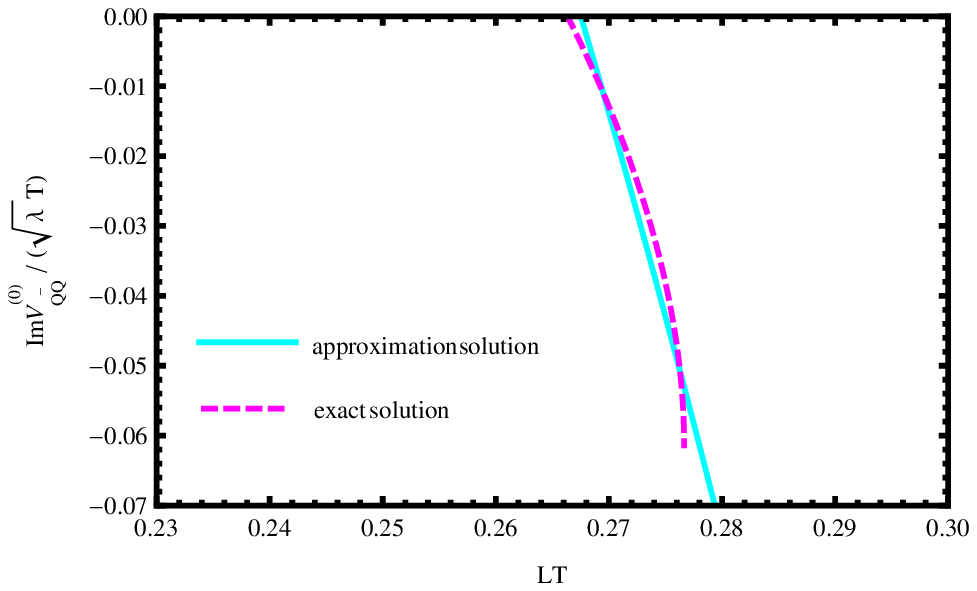}}%
\caption{ Left: The distance between quark antiquark versus the
$\xi$. The specified region in this plot correspond to the valid
region for studying the imaginary potential. Right: The imaginary
potential for the accepted values of $\xi$ or $L$ in the specified
region of right plot. }\label{Lxi0}
\end{figure}

To begin with, we calculate $\Gamma$ in the $N=4$ SYM. The metric
functions are %
\be V(u)=\frac{u^4-u_h^4}{R^4},\,\,\,\,\,W(u)=1.\ee%
where the horizon is located at $u_h$ and the temperature of the hot
plasma is given by $u_h=\pi R^2 T$. In this case we do not consider
any correction and call imaginary
potential as $\text{ImV}_{Q\bar{Q}}^{(0)}$. From \eq{ImV} it is found to be %
\be \text{ImV}_{Q\bar{Q}}^{(0)}= -\frac{\pi \sqrt{\lambda}\,T}{24
\sqrt{2}}\,\left(\frac{3\xi^4-1}{\xi}\right),\,\,\,\,\,\,\,
\xi=\frac{u_h}{u_*}.\label{ImV0}\ee%
The imaginary potential is negative; then it implies that there is a
lower bound for the deepest point of the U-shaped string, i.e.
$\xi_{min}=0.76$. This minimum value
 is found by solving
$\text{ImV}_{Q\bar{Q}}^{(0)}=0$. The maximum value $\xi_{max}$
occurs when the distance $L$ approaches the maximum value. In the
left plot of Fig. \ref{Lxi0}, we show explicitly these values with a
gray filled line.
The imaginary potential for these values of $\xi$ is plotted in the
right plot of Fig.\ref{Lxi0}. It is clearly seen that the imaginary
potential starts to be generated at $L_{min}$ or $\xi_{min}$ and
increases in absolute value with $L$ until a value $L_{max}$ or
$\xi_{max}$. It should be noticed that for $\xi$ very close to the
horizon one should consider higher order corrections
\cite{Bak:2007fk}.

By finding the imaginary potential, the thermal width also is found from \eq{gamma} as %
 \be \Gamma=-\frac{4}{a_0^3}\int_0^\infty L^2 d L e^{-2 L/ a_0}\text{ImV}_{Q\bar{Q}}^{}. \label{TWdefine}\ee
There is a limitation to the calculation of the thermal width from
holography \cite{Fadafan:2013bva}. One finds from Fig. \ref{Lxi0}
that the imaginary potential is well defined for a separation in
$(L_{min}~,~L_{max})$. On the other hand, from a physical point of
view we expect that the imaginary potential should exist also for a
larger separation. One may fix this limitation by assuming that a
solution would be exist at larger distances. We find this solution
by an extrapolating approach. Therefore, we extend the solution in
the $(L_{min}~,~L_{max})$ region to larger distances by
extrapolating the curve. This is the reason why we fitted the
straight line in the right plot of Fig. \ref{Lxi0} which covers
larger distances to infinity.

Briefly, we do the integration in \eq{gamma} in
$(L_{min}~,~L_{max})$ and then we compute the width in the region
$(L_{min}~,~\infty)$ by using a reasonable extrapolation for
imaginary potential. It should be noticed that in
\cite{Noronha:2009da}, the integration in \eq{gamma} has been done
in  $(0~,~\infty)$ and straight line fitting was used for
$L<L_{min}$ region. We do not follow this approach because the
imaginary potential is not defined in $(0~,~L_{min})$. As a result,
in our approach, the value of the width is two orders of magnitude
smaller than the result in [21]. The width in dependence
 on use of the extrapolation method has been discussed also in [24].

We call the width as $\Gamma^{(0)}$.  One finds that for $\Upsilon$
with parameters as $m_Q=4.7GeV, \l=9, R=1$ and at $T=0.3 GeV$ the
width is $ \Gamma^{(0)}=.487 MeV$. In the next sections we will
normalize the thermal widths in the presence of corrections to the $
\Gamma^{(0)}$ value.

\subsection{Thermal width at finite coupling }
From the AdS/CFT, the coupling which is denoted as 't Hooft coupling
$\lambda$ is related to the curvature radius of the $AdS_5$ and
$S_5$ $(R)$, and the string tension $(\frac{1}{2\pi \alpha'})$ by
$\sqrt{\lambda}=\frac{R^2}{\alpha'}$. A general result of the
AdS/CFT correspondence states that the effects of finite but large
$\lambda$ coupling in the boundary field theory are captured by
adding higher derivative interactions in the corresponding
gravitational action. In our study, the corrections to the
AdS-Schwarzschild black brane that will be considered are
$\mathcal{R}^4$ and $\mathcal{R}^2$ corrections. The thermal width
in these cases is called $\Gamma^{(\lambda)}$ and
$\Gamma^{(\lambda_{GB})}$, respectively.

\begin{itemize}%
\item{$\mathcal{R}^4$ corrections:}%
\end{itemize}%
Since the $AdS/CFT$ correspondence refers to the complete string
theory, one should consider the string corrections to the 10D
supergravity action. The first correction occurs at order
$(\alpha')^3$ \cite{alpha2}. In the extremal $AdS_5\times S^5$ it is
clear that the metric does not change \cite{Banks}, conversely this
is no longer true in the non-extremal case. Corrections in inverse
't Hooft coupling $1/\lambda$ which correspond to $\alpha^{\prime}$
corrections on the string theory side, were found in \cite{alpha2}.
The functions of the $\alpha^{\prime}$-corrected metric are given by
\cite{alpha1}
\begin{eqnarray}\label{correctedmetric}
G_{tt}&=&-u^2(1-w^{-4})T(w),\nonumber\\
G_{xx}&=&u^2 X(w),\nonumber\\
G_{uu}&=&u^{-2}(1-w^{-4})^{-1} U(w),
\end{eqnarray}
where%
\begin{eqnarray}
T(w)&=&1-k\bigg(75w^{-4}+\frac{1225}{16}w^{-8}-\frac{695}{16}w^{-12}\bigg)+\dots ,\nonumber\\
X(w)&=&1-\frac{25k}{16}w^{-8}(1+w^{-4})+\dots,\nonumber\\
U(w)&=&1+k\bigg(75w^{-4}+\frac{1175}{16}w^{-8}-\frac{4585}{16}w^{-12}\bigg)+\dots,\
\end{eqnarray}
and $w=\frac{u}{u_h}$. As before, there is an event horizon at
$u=u_h$ and the geometry is asymptotically $AdS$ at large $u$ with a
radius of curvature $R=1$. The expansion parameter $k$ can be
expressed in terms
of the inverse 't Hooft coupling as %
\be\label{k}%
 k=\frac{\zeta(3)}{8}\lambda^{-3/2}\sim 0.15\lambda^{-3/2}.
\ee %
The temperature is given by%
\be %
 T_{}=\frac{u_h}{\pi R^2 (1-k)}.
\ee %

\begin{figure}
\centerline{\includegraphics[width=3in]{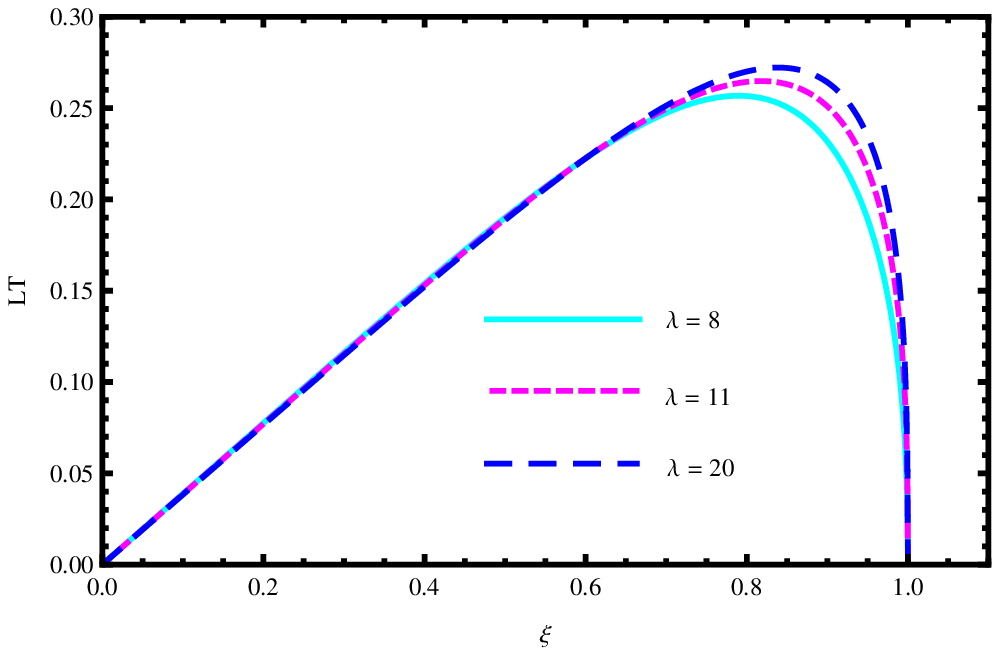}\,\includegraphics[width=3in]{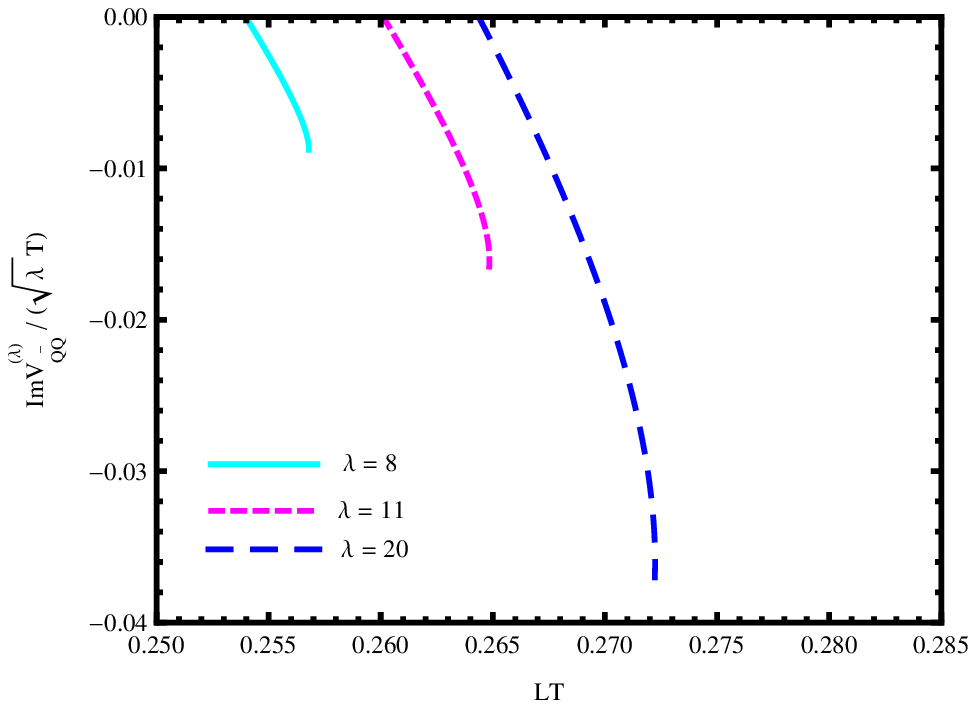}}%
\caption{ Left: The distance between quark-antiquark versus the
$\xi$ for different values of coupling $\lambda_{}$. From top to
down $\lambda_{}$= 20, 11, 8. Right: The imaginary potential for
$\lambda_{}$=8,11,20 from left to right.}\label{LxiR4}
\end{figure}

The imaginary potential is found analytically \eqref{ImV}. However,
in this case it is a lengthy equation. We expand the
result in terms of the expansion parameter $k$ as follows:%
\be
\text{ImV}_{Q\bar{Q}}^{(\lambda)}=\text{ImV}_{Q\bar{Q}}^{(0)}+k\,\frac{\pi
T \sqrt{\l}}{1152 \sqrt{2}}\,g(\xi)+O(k^2) \ee where
$g(\xi)=\left(-48-10656 \xi^4-1775 \xi ^8 + 138840 \xi ^{12}-105495
\xi ^{16}+ 18090 \xi ^{20}\right)/\xi$. The first term is the
imaginary potential in \eqref{ImV0}.

To find $\G^{(\l)}$, we study behavior of the quark-antiquark
distance in terms of $\xi$ for different values of $\lambda$. As
shown in the left plot of Fig. \ref{LxiR4} by increasing the
coupling constant the maximum value of $L$ increases. In the right
plot of this figure we show effect of coupling on the imaginary
potential. One finds that by increasing the coupling, $L_{min}$ also
increases. As an example, by choosing $R=1$, $T=0.3GeV$ and for
$\lambda=8,11,20$, we find $L_{min}T \simeq 0.254, 0.260, 0.264$ and
$L_{max}T \simeq 0.257, 0.265, 0.272$. We conclude that the
imaginary part of the potential in the presence of coupling
corrections is
generated for larger distances:%
\be L_{min}^{\l_1}<L_{min}^{\l_2}<L_{min}^{\infty}. \ee%
where $\l_1 < \l_2< \infty$. Therefore, we find that in absolute
value the imaginary potential is increased due to finite coupling
corrections.

\begin{figure}
\centerline{\includegraphics[width=3in]{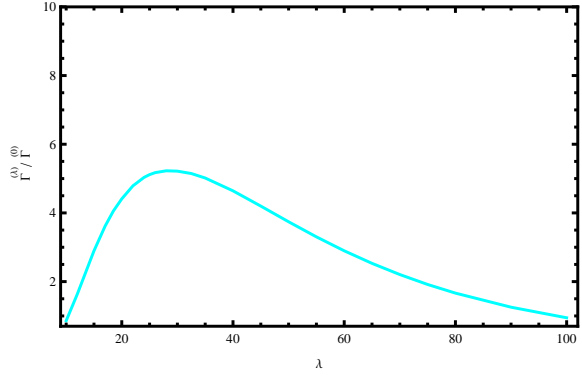}}%
\caption{The thermal width versus the finite coupling correction
computed by considering $\mathcal{R}^4$ corrections. We used the
gauge theory coupling $\l =100$ in this case. }\label{TWR4}
\end{figure}

It would be important to notice that we cannot use the extrapolation
method in this case. This is so because the sign of the imaginary
potential for $\xi > \xi_{max}$ is not always negative and changes.
Then we cannot use a straight line approximation to consider the
larger values of $L$ like we did in the right plot of Fig.
\ref{Lxi0}. Using this point we integrate the thermal width
$\Gamma^{(\lambda)}$ from $L_{min}$ to $L_{max}$. We show
$\Gamma^{(\lambda)}$ relative to the $\Gamma^{(0)}$ in this interval
in Fig. \ref{TWR4}. We used the gauge theory coupling $\l =100$ in
this case to calculate $\Gamma^{(0)}$.  As is clear from this
figure, when $\lambda$ goes to infinity
$\frac{\Gamma^{(\lambda)}}{\Gamma^{(0)}}$ goes to unity. The maximum
value of $\frac{\Gamma^{(\lambda)}}{\Gamma^{(0)}}$ equals to $5.22$,
which occurs at $\l_c=28$.

At strong coupling, an estimate of how the thermal width of heavy
quarkonium changes with the shear viscosity to entropy density
ratio, $\eta/s$, is studied in \cite{Finazzo:2013rqy}. It is found
that in the presence of curvature-squared corrections like
Gauss-Bonnet terms, the thermal width as a function of $\eta/s$
decreases. It has been shown that considering $\mathcal{R}^4$
corrections increases $\eta/s$ \cite{Buchel:2004di}. One concludes
that decreasing $\l$ leads to increasing $\eta/s$. Now, one finds
from Fig. \ref{TWR4} that by decreasing $\l$ from $\l_c$, which
means increasing $\eta/s$, the width becomes effectively smaller.
Interestingly this is similar to analogous calculations in
perturbative QCD \cite{First}. In this study, it was argued that at
strong coupling for a quarkonium with a very heavy constituent mass,
the thermal width can be ignored \cite{First} which is in good
agreement with our result.

\begin{itemize}%
\item{$\mathcal{R}^2$ corrections:}%
\end{itemize}%
Next, we study $\mathcal{R}^2$ corrections to the thermal width
which is called $\Gamma^{(\lambda_{GB})}$.

In five dimensions, we consider the theory of gravity with quadratic
powers of curvature as Gauss-Bonnet(GB) theory. The exact solutions
and thermodynamic properties of the black brane in GB gravity are
discussed in \cite{Cai:2001dz,Nojiri:2001aj,Nojiri:2002qn}. The metric functions are given by%
\begin{equation}
G_{tt}=-N \,u^2\, h(u),\,\,\,\,\,\, G_{uu}=\frac{1}{u^2
h(u)},\,\,\,\,\, G_{xx}=G_{yy}=G_{zz}=u^2\label{GBmetric},
\end{equation}
where
\begin{equation}
h(u)= \frac{1}{2\lambda_{GB}}\left[ 1-\sqrt{1-4 \lambda_{GB}\left(
1-\frac{u_h^4}{u^4} \right)}\right].
\end{equation}
In (\ref{GBmetric}), $N= \frac{1}{2}\left(
 1+\sqrt{1-4 \lambda_{GB}} \right)$ which is an arbitrary constant that specifies
the speed of light of the boundary gauge theory and we choose it to
be unity. Beyond $\l_{GB}<1/4$ there is no vacuum AdS solution and
one cannot have a conformal field theory at the boundary. Causality
leads to new bounds for the value of the Gauss-Bonnet coupling
constant as follows: $-7/36<\l_{GB}<9/100$ \cite{Brigante008gz}. The
temperature also is given by
\begin{equation}
 T_{}=\sqrt{N}\,\frac{u_h}{\pi R^2}.
\end{equation}
Also, the 't Hooft coupling of the dual strongly coupled CFT is
$\l=\frac{N^2 R^4}{\a'^2}$ .
\begin{figure}
\centerline{\includegraphics[width=3in]{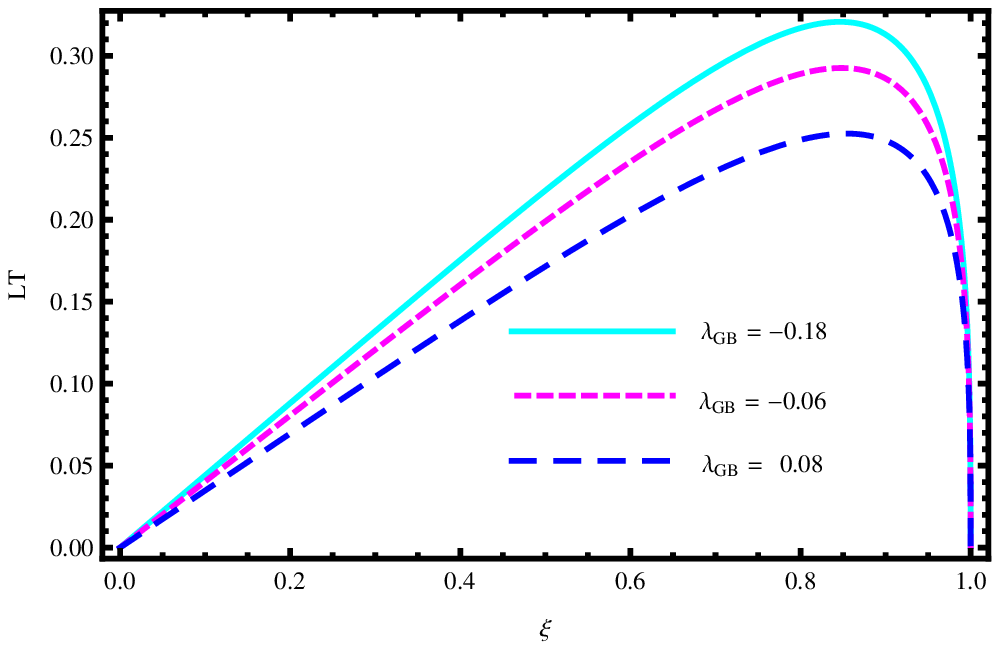}\,\,\includegraphics[width=3in]{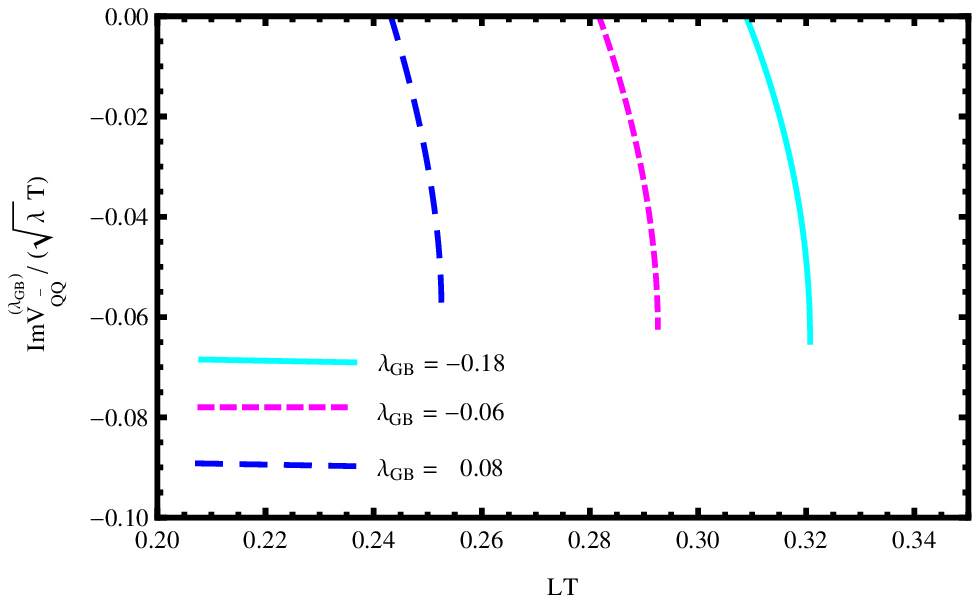}}%
\caption{ Left: The distance between quark-antiquark versus the
$\xi$ for different values of Gauss-Bonnet coupling $\lambda_{GB}$.
From top to down $\lambda_{GB}$=-0.18, -0.06, 0.08. Right: the
imaginary potential versus the Gauss-Bonnet coupling constant. From
right to left $\lambda_{GB}$=-0.18, -0.06, 0.08. }\label{LxiGB}
\end{figure}

The thermal width in this background has been studied in
\cite{Finazzo:2013rqy}. Also the behavior of the width in terms of
the shear viscosity to entropy density ratio was found. Here, we
present further details and express the final results in terms of
the Gauss-Bonnet coupling constant. As it was pointed out, our
extrapolation method is different from \cite{Finazzo:2013rqy}.

The behavior of $L$ in terms of $\xi$ for different values of
$\lambda_{GB}$ is shown in Fig. \ref{LxiGB}. It is clearly seen in
the left plot of Fig. \ref{LxiGB} that by increasing the coupling
constant the maximum value of $L$ decreases. This behavior is
not the same as the $\mathcal{R}^4$ corrections in Fig.\ref{LxiR4}.%

The imaginary potential in the Gauss-Bonnet gravity results from
\eqref{ImV}. The result can be expressed as follows:%
\bea &&\text{ImV}_{Q\bar{Q}}^{(\lambda_{GB})}=-\frac{h \pi
T\sqrt{\lambda }}{4 \sqrt{2} N  \xi} \\&&\frac{ 1-h+\left(-8+6
h+(9-7 h) \xi ^4\right) \lambda _{\text{GB}}+4 \left(4-2 h+(-9+5 h)
\xi ^4+3 \xi ^8\right) \lambda _{\text{GB}}^2} { \left(1-h+2
\left(-2+\xi ^4\right) \lambda _{\text{GB}}\right) \left(3-3 h-6
\left(4-2 h+(-3+2 h) \xi ^4\right) \lambda _{\text{GB}}+
8 \left(6-9 \xi ^4+\xi ^8\right) \lambda _{\text{GB}}^2\right)},\nn\eea%
where \be h=\sqrt{1+4 \left(-1+\xi ^4\right) \text{$\l_{GB}$}}.\ee

In the right plot of this figure we show effect of coupling on the
imaginary potential. For example, by choosing $R=1$, $T=0.3GeV$,
$\l=9$ and for $\l_{GB}=-0.18,-0.06,0.08$, we find $L_{min}T \simeq
0.30, 0.28,0.24$ and $L_{max}T \simeq 0.32,0.29,0.25$. Comparing
with $\mathcal{R}^4$ corrections, one finds different behavior, i.e.
by increasing the coupling $L_{min}$ decreases. This means that the
imaginary potential in the Gauss-Bonnet gravity starts to be
generated for smaller distances. We find that the absolute value of
the imaginary part of the potential is increased due to finite
coupling corrections.

\begin{figure}
\centerline{,\includegraphics[width=3in]{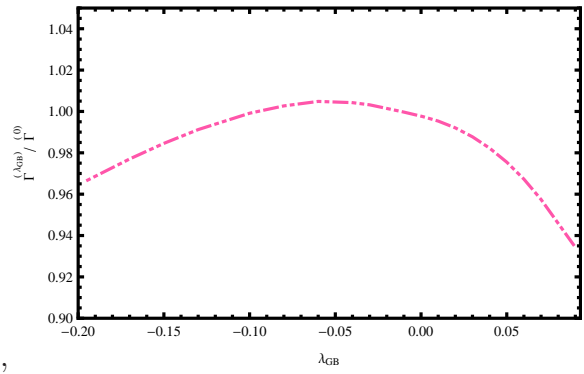}}%
\caption{The normalized thermal width versus the Gauss-Bonnet
coupling correction. There is a maximum value at $\l_{GB}=-0.06$.}
\label{TWGB}
\end{figure}

In this case also we should do the integration in \eqref{TWdefine}
in $(L_{min}~,~L_{max})$ and one cannot consider larger lengths by
extrapolating the curves in Fig.\ref{LxiGB}. We would like to
emphasize that using this method leads to the Fig.\ref{TWGB}. The
modifications due to the finite coupling corrections to the
Coulombic part of the real potential have been considered, too. At
$\lambda_{GB}=-0.06$, there is a maximum value for
$\G^{(GB)}/\G^{(0)}$ which is $1.004$. For other values of the
coupling constant, the thermal width decreases.

\subsection{Medium effect on the thermal width }
In this section, we consider the quarkonium in the medium composed
of light quarks and gluons \cite{Kim:2007em}. It is shown that at
the high temperature, the gravity dual to the QGP phase is the
Reissner-Nordstrom AdS $(RNAdS)$ black hole and at the low
temperature, the dual geometry to the hadronic phase is the thermal
charged AdS $(tcAdS)$ space. The confinement/deconfinement phase
transition in the quark medium is discussed in \cite{Kim:2007em} and
an influence of matters on the deconfinement temperature, $T_c$, is
investigated. Using a different normalization for the bulk gauge
field, it is shown that the critical baryonic chemical potential
becomes $1100 MeV$ which is comparable to the QCD result
\cite{Park:2009nb}. Melting of a heavy meson is investigated in
\cite{Fadafan:2012qy} and it is found that the melting mechanism in
the QGP and in the hadronic phase are the same, i.e. the interaction
between heavy quarks is screened by the light quarks. The drag force
on a moving heavy quark and the jet quenching parameter in the
background of RNAdS black hole are studied in \cite{Fadafan:2008uv}.

Before calculating the thermal width, we will give a brief review of
\cite{Kim:2007em}. In this background the density in the dual field
theory is mapped to a bulk gauge field. The Euclidean action
describing the five-dimensional asymptotic $
AdS$ space with the gauge field is given by%

\be S = \int d^5 x \sqrt{G} \left( \frac{1}{2 \kappa^2} \left( -
{\cal R} + 2 \Lambda\right)  + \frac{1}{4g^2} F_{MN} F^{MN} \right)
. \label{S1}\ee%
Here $\kappa^2$ is proportional to the five-dimensional Newton
constant and $g^2$ is a five-dimensional gauge coupling constant.
The cosmological constant is given by $\Lambda = \frac{-6}{R^2}$,
where $R$ is the radius of the $AdS$ space.

As it was pointed out the QGP and hadronic phases are described by
the $tcAdS$ and the $RNAdS$ black hole, respectively
\cite{Park:2009nb}. It was argued in section 2 that the imaginary
part of the potential in the confinement phase would be zero, then
we focus on the QGP phase. This solution is considered as follows:%

\be ds^2= R^2 u^2 \left( - f(u) dt^2 + d \vec{x}^{ 2} + \frac{1}{u^4
f(u)} du^2 \right)  ,
\label{ds2}\ee%
where the blackness function $f(u)$ is given by%

\be f(u)=1-m u^{-4}+ q^2 u^{-6}.\label{fQGP}\ee%
In these coordinates, $u$ denotes the radial coordinate of the black
hole geometry and $t, \vec{x}$ label the directions along the
boundary at the spatial infinity. The boundary is located at
infinity and the geometry is asymptotically $AdS$ with radius $R$.
The event horizon is located at $f(u_h)=0$ where $u_h$ is the
largest root of this equation. The black hole mass $(m)$ and the
temperature $(T)$ are given by %
 \be m=u_h^4+q^2u_h^{-2},\,\,\,\,\,\,T=\frac{u_h}{\pi} \left( 1- \frac{q^2}{2 u_h^6}\right).\ee
where $q$ is the black hole charge.
\begin{figure}
\centerline{\includegraphics[width=3in]{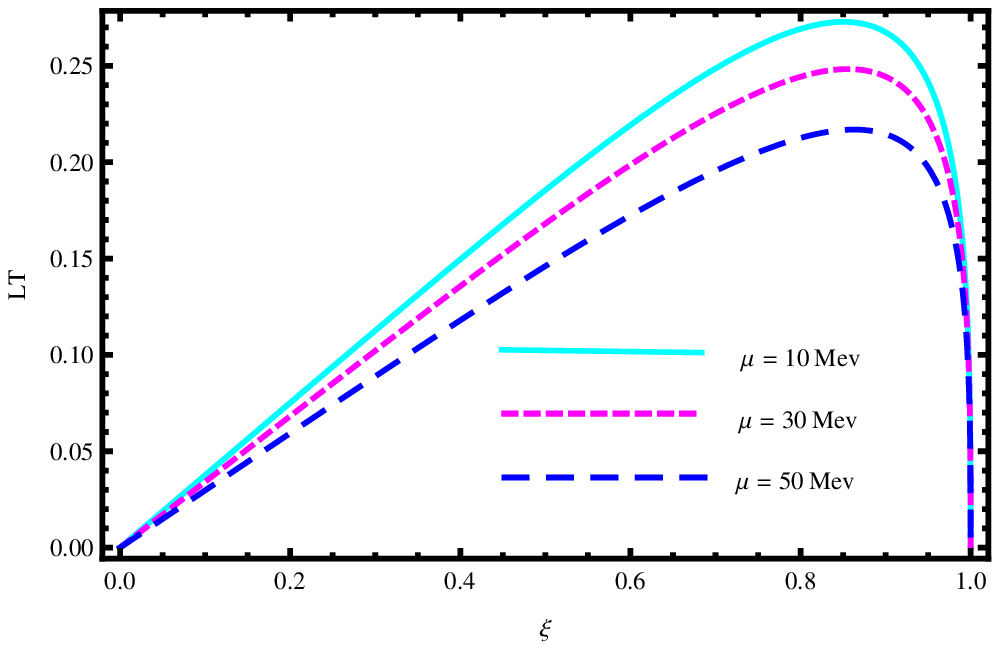}\includegraphics[width=3in]{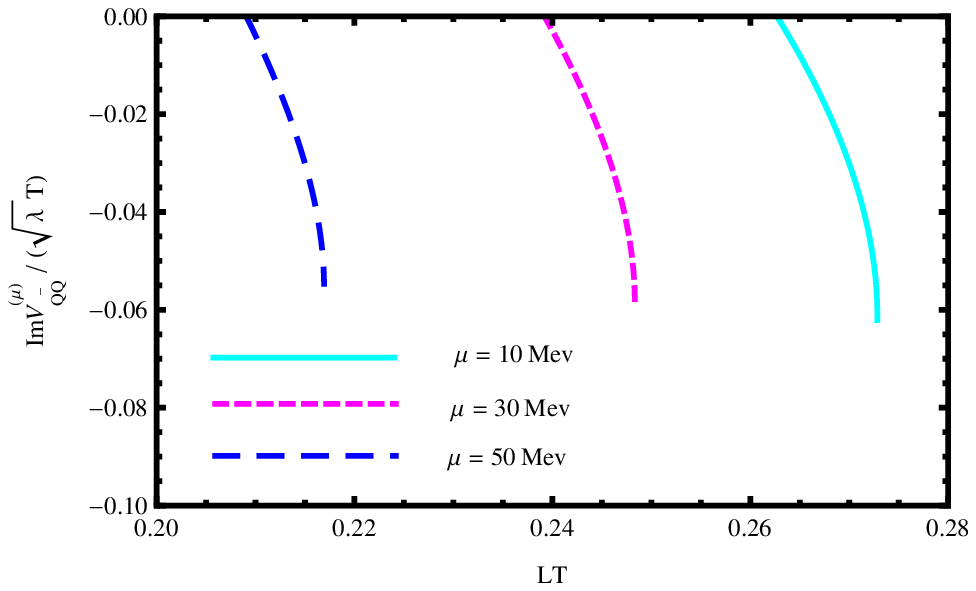}}%
\caption{ Left: The distance between quark-antiquark versus the
$\xi$ for different values of chemical potential $\mu$. From top to
down $\mu$=10MeV, 30MeV, 50MeV .Right: The imaginary potential
versus the $L$ for $\mu$=10MeV, 30MeV, 50MeV .}\label{Lximu}
\end{figure}

The time-component of the bulk gauge field is
$A_t(u)=i(2\pi^2\mu-Q\,u^{-2})$ where $\mu$ and $Q$ are related to
the chemical potential and quark number density in the dual gauge
theory. Regarding the Drichlet boundary condition at the horizon,
$A_t(u_h)=0$, one finds $Q=2 \pi^2 \mu \,u_h$. The black hole charge
$q$ and the quark number density $Q$
are also related to each other by this equation%
\be Q=\sqrt{\frac{3}{2n}}\,q. \ee%
where $n^{-1}=\frac{g^2R^2}{\kappa^2}$ is the color number.

Now using the general results in (3.4) and \eqref{staticV}, we study
the behavior of $L$ in terms of $\xi$ for different values of $\mu$.
The results are shown in Fig. \ref{Lximu}. It is clearly seen in the
left plot of this figure that by increasing the chemical potential
the maximum value of $L$ decreases. On the contrary, considering
$\mathcal{R}^4$ corrections increases the maximum value of $L$ which
can be seen in Fig. \ref{LxiR4}.

The imaginary potential in this background is found from
\eqref{ImV}. The result is expressed as follows:%
\be \text{ImV}_{Q\bar{Q}}^{(\mu)}=\frac{\sqrt{\lambda } u_h \left(32 n^2 \pi ^8 \mu ^4 \xi ^{10}
\left(3-2 \xi ^2\right)+12 n \pi ^4 \mu ^2 \xi ^4 \left(6-13 \xi ^2+3 \xi ^6\right)
u_h^2+9 \left(-1+3 \xi ^4\right) u_h^4\right)}{24 \sqrt{2} \left(16 n^2
\pi ^8 \mu ^4 \xi ^{13}-9 \xi  u_h^4\right)}.\ee%
Here at $\mu=0$, one finds the exact result in \eqref{ImV0}.

In the right plot of Fig. \ref{Lximu}, we show effect of chemical
potential on the imaginary potential. We fixed the parameters as
$R=1$, $\l=9$, $T=0.3GeV$ and $n=1$. For $\mu=10MeV, 30 MeV$ and
$\mu=50MeV$, one finds $L_{min}T \simeq 0.26, 0.23, 0.20$ and
$L_{max}T \simeq 0.27,0.24, 0.21$. Comparing with $\mathcal{R}^4$
corrections, one finds different behavior, i.e. by increasing the
chemical potential $L_{min}$ decreases:%
\be L_{min}^{\m_1}<L_{min}^{\m_2}<L_{min}^{\m_3} \ee%
where $\m_1 > \m_2 > \m_3$. As a result, one concludes that the
imaginary potential in the presence of light quarks starts to be
generated for smaller distances, while at finite $\l$ coupling it
starts at a larger distance.
\begin{figure}
\centerline{\includegraphics[width=3in]{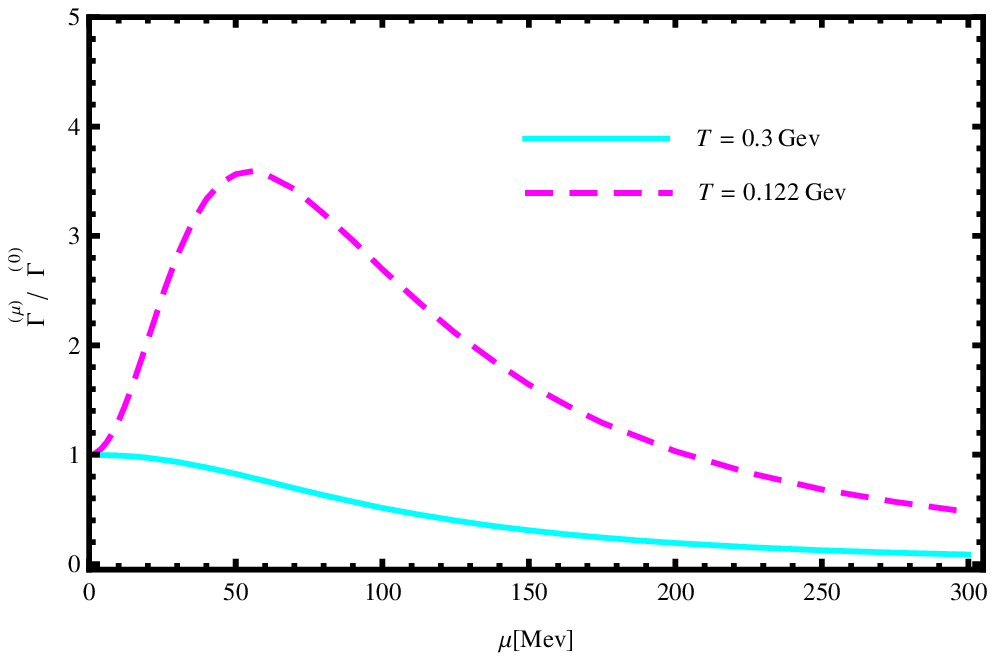}\includegraphics[width=3in]{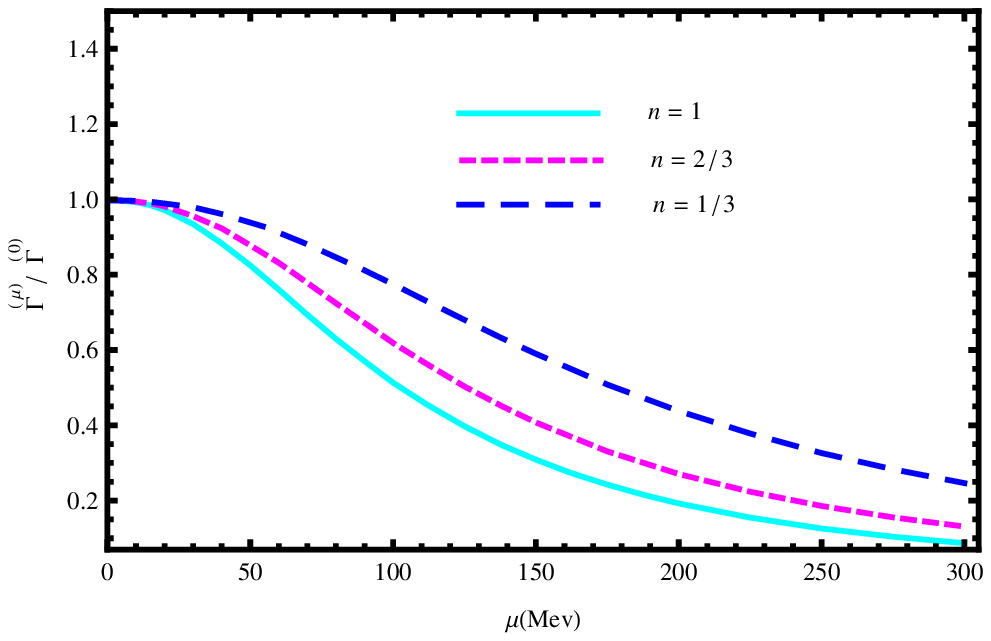}}%
\caption{Thermal width in the presence of light quarks. Left: The
ratio of $\frac{\G^{(\m)}}{\G^{(0)}}$ is shown versus the chemical
potential. Two different temperatures from top to down are $T=0.122
GeV$ and $T=0.3 GeV$. Right: The effect of increasing flavor color
number on the ratio of $\frac{\G^{(\m)}}{\G^{(0)}}$ at $T=0.3 GeV$.
}\label{TWmu}
\end{figure}

To calculate the thermal width, one should do the integration in
\eqref{TWdefine} in $(L_{min}~,~L_{max})$.  It is found that by
increasing the temperature, $\G^{(\m)}$ also increases. At fixed
temperature, there is a maximum value for $\G^{(\m)}$. The ratio of
$\frac{\G^{(\m)}}{\G^{(0)}}$ versus the chemical potential is shown
in the Fig. \ref{TWmu}. In this figure, two different temperatures
from top to down are $T=0.122 GeV$ and $T=0.3 GeV$, respectively.

We introduce  $N_f = n N_c$ where $N_f$ and $N_c$ are the number of
flavors and color fields, respectively. Now one can find the
behavior of the thermal width when the flavor number of quarks
increases. In the right plot of Fig.\ref{TWmu}, the effect of
increasing flavor color number on the ratio of
$\frac{\G^{(\m)}}{\G^{(0)}}$ has been shown at $T=0.3 GeV$. It is
seen that the maximum value of thermal width does not change but the
rate of it versus the chemical potential changes, monotonically.

\section{Conclusion}
In this paper, we studied the different effects on the thermal width
from holography. We considered the effects of the charge and the
finite 't Hooft coupling correction on the hot plasma. An
understanding of how the width changes by these corrections may be
essential for theoretical predictions in perturbative QCD
\cite{Brambilla:2013dpa,Brambilla:2011sg}. To study the charge
effect, we considered the Maxwell charge, which is interpreted as a
quark medium. The effects of finite but large couplings are
considered by adding higher derivative corrections in the gravity
background. Especially, $\mathcal{R}^4$ terms and Gauss-Bonnet
gravity have been studied.

As it was found in \cite{Fadafan:2013bva} the minimum distance of
the quark-antiquark pair, $L_{min}$, where the imaginary potential
starts depends on the corrections. Our findings in this case can be
summarized as follows:%
\begin{itemize}%
\item{In the presence of $\mathcal{R}^4$ corrections which correspond to finite 't Hooft coupling
corrections in the hot plasma, by increasing the coupling $L_{min}$ also increases.  }%
\item{By considering Gauss-Bonnet corrections, increasing the Gauss-Bonnet coupling leads to decreasing of $L_{min}$.}%
\item{In the medium, increasing $\mu$ leads to decreasing $L_{min}$. }
\end{itemize}%
We normalized the thermal width of quarkonium  to $\G^{(0)}$ which is $N=4$ SYM result. The following is shown.%
\begin{itemize}%
\item{ By turning on the 't Hooft coupling
correction in the hot plasma, $\frac{\G^{(\l)}}{\G^{(0)}}$ increases
up to a maximum value. Decreasing of
 the coupling leads to smaller width. One concludes that at finite 't Hooft coupling the width becomes smaller.
}%
\item{There is a maximum value for $\frac{\G^{(GB)}}{\G^{(0)}}$ in terms of $\l_{GB}$.}%
\item{In the presence of a medium $\frac{\G^{(\m)}}{\G^{(0)}}$ takes a maximum
value which depends on the temperature of the hot plasma. The effect
of flavor number is shown in the right plot of Fig.\ref{TWmu}.  }
\end{itemize}%
In the case of $\mathcal{R}^4$ corrections, the relation between the
thermal width of heavy quarkonium and the shear viscosity to entropy
density ratio, $\eta/s$, was discussed. It was found that in the
presence of these corrections, by decreasing $\l$, which means
increasing $\eta/s$, the thermal width becomes effectively smaller.
This is an interesting result which is consistent with the intuition
one would get from weakly coupled plasma \cite{First}.

It will be very interesting to investigate the thermal width of a
quarkonia in more realistic holographic backgrounds, such as in
\cite{G1} and \cite{Lee:2013oya}. Finally, comparing holographic
results with weakly coupled calculations would be desirable. We
would like to report on this study elsewhere.

\section*{Acknowledgment}
The authors thank M. Ali-Akbari, E. Azimfard, H. Soltanpanahi and D.
Giataganas for very useful discussions.  We are very grateful to and
thank N. Brambilla for a discussion of the related papers in the
subject of weakly coupled plasma. We would like also to thank J.
Noronha for important comments on this manuscript and M. Sohani for
reading it, carefully.




\begin{thebibliography}{99}
\bibitem{CasalderreySolana:2011us}
  J.~Casalderrey-Solana, H.~Liu, D.~Mateos, K.~Rajagopal and U.~A.~Wiedemann,
  ``Gauge/String Duality, Hot QCD and Heavy Ion Collisions,''
  arXiv:1101.0618 [hep-th].

\bibitem{ALICE}
The ALICE Collaboration, K. Aamodt et al., "Elliptic fow of charged
particles in Pb-Pb collisions at 2.76 TeV," arXiv:1011.3914
[nucl-ex].

\bibitem{Maldacena:1997re}
  J.~M.~Maldacena,
  ``The large N limit of superconformal field theories and supergravity,''
  Adv.\ Theor.\ Math.\ Phys.\  {\bf 2} (1998) 231
  [Int.\ J.\ Theor.\ Phys.\  {\bf 38} (1999) 1113]
  [arXiv:hep-th/9711200].

\bibitem{Gubser:1998bc}
  S.~S.~Gubser, I.~R.~Klebanov and A.~M.~Polyakov,
  ``Gauge theory correlators from non-critical string theory,''
  Phys.\ Lett.\  B {\bf 428} (1998) 105
  [arXiv:hep-th/9802109].
\bibitem{Witten:1998qj}
  E.~Witten,
  ``Anti-de Sitter space and holography,''
  Adv.\ Theor.\ Math.\ Phys.\  {\bf 2} (1998) 253
  [arXiv:hep-th/9802150].
\bibitem{Witten:1998zw}
  E.~Witten,
  ``Anti-de Sitter space, thermal phase transition, and confinement in  gauge
  theories,''
  Adv.\ Theor.\ Math.\ Phys.\  {\bf 2} (1998) 505
  [arXiv:hep-th/9803131].
\bibitem{DeWolfe:2013cua}
  O.~DeWolfe, S.~S.~Gubser, C.~Rosen and D.~Teaney,
  ``Heavy ions and string theory,''  arXiv:1304.7794 [hep-th].


\bibitem{Laine:2011xr}
  M.~Laine,
  ``News on hadrons in a hot medium,''
  arXiv:1108.5965 [hep-ph].
\bibitem{Matsui:1986dk}
  T.~Matsui and H.~Satz,
  ``J/psi Suppression by Quark-Gluon Plasma Formation,''
  Phys.\ Lett.\ B {\bf 178} (1986) 416.
\bibitem{Brambilla:2004wf}
  N.~Brambilla {\it et al.}  [Quarkonium Working Group Collaboration],
  ``Heavy quarkonium physics,''  hep-ph/0412158.\\
  N.~Brambilla, S.~Eidelman, B.~K.~Heltsley, R.~Vogt, G.~T.~Bodwin,
   E.~Eichten, A.~D.~Frawley and A.~B.~Meyer {\it et al.},
  ``Heavy quarkonium: progress, puzzles, and opportunities,
  ''  Eur.\ Phys.\ J.\ C {\bf 71} (2011) 1534  [arXiv:1010.5827 [hep-ph]].
\bibitem{First}
M. Laine, O. Philipsen, P. Romatschke, and M. Tassler, "Real-time
static potential in hot QCD," JHEP 03 (2007) 054,
arXiv:hep-ph/0611300.\\
M. Laine, " resummed perturbative estimate for the quarkonium
spectral function in hot QCD," JHEP 05 (2007) 028, arXiv:0704.1720
[hep-ph].

\bibitem{Beraudo:2007ky}
  A.~Beraudo, J.~-P.~Blaizot and C.~Ratti,
  ``Real and imaginary-time Q anti-Q correlators in a thermal medium,
  ''  Nucl.\ Phys.\ A {\bf 806} (2008) 312  [arXiv:0712.4394 [nucl-th]].

\bibitem{Brambilla:2008cx}
  N.~Brambilla, J.~Ghiglieri, A.~Vairo and P.~Petreczky,
  ``Static quark-antiquark pairs at finite temperature,''
   Phys.\ Rev.\ D {\bf 78} (2008) 014017  [arXiv:0804.0993 [hep-ph]].
\bibitem{Brambilla:2010vq}
  N.~Brambilla, M.~A.~Escobedo, J.~Ghiglieri, J.~Soto and A.~Vairo,
  ``Heavy Quarkonium in a weakly-coupled quark-gluon plasma below the melting temperature,''
    JHEP {\bf 1009} (2010) 038  [arXiv:1007.4156 [hep-ph]].
\bibitem{Brambilla:2011sg}
  N.~Brambilla, M.~A.~Escobedo, J.~Ghiglieri and A.~Vairo,
  ``Thermal width and gluo-dissociation of quarkonium in pNRQCD,''
   JHEP {\bf 1112} (2011) 116  [arXiv:1109.5826 [hep-ph]].
\bibitem{Dumitru:2010id}
  A.~Dumitru,
  ``Quarkonium in a non-ideal hot QCD Plasma,
  ''  Prog.\ Theor.\ Phys.\ Suppl.\  {\bf 187} (2011) 87  [arXiv:1010.5218 [hep-ph]].
\bibitem{Rothkopf:2011db}
  A.~Rothkopf, T.~Hatsuda and S.~Sasaki,
``Complex Heavy-Quark Potential at Finite Temperature from Lattice
QCD,
  ''  Phys.\ Rev.\ Lett.\  {\bf 108} (2012) 162001  [arXiv:1108.1579 [hep-lat]].
\bibitem{Fujita:2009wc}
  M.~Fujita, K.~Fukushima, T.~Misumi and M.~Murata,
  ``Finite-temperature spectral function of the vector mesons in an AdS/QCD model,
  ''  Phys.\ Rev.\ D {\bf 80} (2009) 035001  [arXiv:0903.2316 [hep-ph]].


\bibitem{imaginary2} M. Margotta, K. McCarty, C. McGahan, M. Strickland,
and D. Yager-Elorriaga, "Quarkonium states in a complex-valued
potential," Phys.Rev. D83 (2011) 105019, arXiv:1101.4651 [hep-ph].
\bibitem{Margotta:2011ta}
  M.~Margotta, K.~McCarty, C.~McGahan, M.~Strickland and D.~Yager-Elorriaga,
  ``Quarkonium states in a complex-valued potential,''  Phys.\ Rev.\ D {\bf 83} (2011) 105019
     [Erratum-ibid.\ D {\bf 84} (2011) 069902]  [arXiv:1101.4651 [hep-ph]].


\bibitem{Noronha:2009da}
  J.~Noronha and A.~Dumitru,
  ``Thermal Width of the $\Upsilon$ at Large t' Hooft Coupling,''  Phys.\ Rev.\ Lett.\  {\bf 103} (2009) 152304  [arXiv:0907.3062 [hep-ph]].
\bibitem{Dimitris1}
  D.~Giataganas,
  ``Observables in Strongly Coupled Anisotropic Theories,''  arXiv:1306.1404 [hep-th].
\bibitem{Fadafan:2013bva}
  K.~B.~Fadafan, D.~Giataganas and H.~Soltanpanahi,
  ``The Imaginary Part of the Static Potential in Strongly Coupled Anisotropic Plasma,''
  arXiv:1306.2929 [hep-th].
\bibitem{Finazzo:2013rqy}
  S.~I.~Finazzo and J.~Noronha,
  ``Estimates for the Thermal Width of Heavy Quarkonia in Strongly Coupled Plasmas from Holography,''
  arXiv:1306.2613 [hep-ph].



\bibitem{Erdmenger:2007cm}
  J.~Erdmenger, N.~Evans, I.~Kirsch and E.~Threlfall,
 ``Mesons in Gauge/Gravity Duals - A Review,''
 Eur.\ Phys.\ J.\ A {\bf 35} (2008) 81
 [arXiv:0711.4467 [hep-th]]..
\bibitem{Albacete:2008dz}
  J.~L.~Albacete, Y.~V.~Kovchegov and A.~Taliotis,
  ``Heavy Quark Potential at Finite Temperature in AdS/CFT Revisited,''
  Phys.\ Rev.\ D {\bf 78} (2008) 115007
  [arXiv:0807.4747 [hep-th]].
\bibitem{Hayata:2012rw}
  T.~Hayata, K.~Nawa and T.~Hatsuda,
  ``Time-dependent Heavy-Quark Potential at Finite Temperature from Gauge/Gravity Duality,''
  arXiv:1211.4942 [hep-ph].



\bibitem{Bak:2007fk}
  D.~Bak, A.~Karch and L.~G.~Yaffe,
  ``Debye screening in strongly coupled N=4 supersymmetric Yang-Mills plasma,''
  JHEP {\bf 0708} (2007) 049
  [arXiv:0705.0994 [hep-th]].



\bibitem{Maldacena:1998im}
  J.~M.~Maldacena,
  ``Wilson loops in large N field theories,''
  Phys.\ Rev.\ Lett.\  {\bf 80} (1998) 4859
  [hep-th/9803002].
\bibitem{Rey}
  S.~-J.~Rey, S.~Theisen and J.~-T.~Yee,
  ``Wilson-Polyakov loop at finite temperature in large N gauge theory and anti-de Sitter supergravity,''  Nucl.\ Phys.\ B {\bf 527} (1998) 171  [hep-th/9803135].  
  A.~Brandhuber, N.~Itzhaki, J.~Sonnenschein and S.~Yankielowicz,
  ``Wilson loops in the large N limit at finite temperature,''  Phys.\ Lett.\ B {\bf 434} (1998) 36  [hep-th/9803137].  
\bibitem{Noronha:2009ia}
  J.~Noronha and A.~Dumitru,
  ``The Heavy Quark Potential as a Function
  of Shear Viscosity at Strong Coupling,''  Phys.\ Rev.\ D {\bf 80} (2009) 014007  [arXiv:0903.2804 [hep-ph]].
\bibitem{Fadafan:2011gm}
  K.~B.~Fadafan,
  ``Heavy quarks in the presence of higher derivative corrections from AdS/CFT,''
  Eur.\ Phys.\ J.\ C {\bf 71} (2011) 1799
  [arXiv:1102.2289 [hep-th]].
\bibitem{Peeters}
  K.~Peeters, J.~Sonnenschein and M.~Zamaklar,
  ``Holographic melting and related properties of mesons in a quark gluon
  plasma,''
  Phys.\ Rev.\  D {\bf 74} (2006) 106008
  [arXiv:hep-th/0606195];
\bibitem{Antipin}
  O.~Antipin, P.~Burikham and J.~Li,
  ``Effective Quark Antiquark Potential in the Quark Gluon Plasma from
  Gravity Dual Models,''
  JHEP {\bf 0706} (2007) 046
  [arXiv:hep-ph/0703105];
  P.~Burikham and J.~Li,
  ``Aspects of the screening length and drag force in two alternative gravity
 duals of the quark-gluon plasma,''
  JHEP {\bf 0703}, 067 (2007)
  [arXiv:hep-ph/0701259];

\bibitem{AliAkbari:2009pf}
  M.~Ali-Akbari and K.~Bitaghsir Fadafan,
  ``Rotating mesons in the presence of higher derivative corrections from
  gauge-string duality,''
  Nucl.\ Phys.\  B {\bf 835} (2010) 221
  [arXiv:0908.3921 [hep-th]].

\bibitem{Brambilla:2013dpa}
  N.~Brambilla, M.~A.~Escobedo, J.~Ghiglieri and A.~Vairo,
  ``Thermal width and quarkonium dissociation by inelastic parton scattering,''
  JHEP {\bf 1305} (2013) 130  [arXiv:1303.6097 [hep-ph]].
\bibitem{alpha2}
 J. Pawelczyk and S. Theisen,
 {\it AdS$_5\times S^5$ black hole metric at {\cal O}($\alpha^{\prime 3}$)},
  JHEP {\bf 9809} (1998) 010, [hep-th/9808126];

\bibitem{Banks}
  T.~Banks and M.~B.~Green,
  ``Non-perturbative effects in AdS(5) x S**5 string theory and d = 4 SUSY
  Yang-Mills,''
  JHEP {\bf 9805}, 002 (1998)
  [arXiv:hep-th/9804170];

\bibitem{alpha1}
 S.S. Gubser, I.R. Klebanov and A.A. Tseytlin,
{ \it Coupling constant dependence in the thermodynamics of $N=4$
supersymmetric Yang-Mills theory}
 Nucl. Phys. {\bf B534} (1998) 202, [hep-th/9805156];
\bibitem{Buchel:2004di}
  A.~Buchel, J.~T.~Liu and A.~O.~Starinets,
  ``Coupling constant dependence of the shear viscosity in
  N=4 supersymmetric Yang-Mills theory,''  Nucl.\ Phys.\ B {\bf 707} (2005) 56  [hep-th/0406264].
\bibitem{Cai:2001dz}
  R.~G.~Cai,
  ``Gauss-Bonnet black holes in AdS spaces,''
  Phys.\ Rev.\  D {\bf 65} (2002) 084014
  [arXiv:hep-th/0109133].
\bibitem{Nojiri:2001aj}
  S.~Nojiri and S.~D.~Odintsov,
  ``Anti-de Sitter black hole thermodynamics in higher derivative gravity  and
  new confining-deconfining phases in dual CFT,''
  Phys.\ Lett.\  B {\bf 521} (2001) 87
  [Erratum-ibid.\  B {\bf 542} (2002) 301]
  [arXiv:hep-th/0109122].
\bibitem{Nojiri:2002qn}
  S.~Nojiri and S.~D.~Odintsov,
  "(Anti-) de Sitter black holes in higher derivative gravity and dual
  conformal field theories,"
  Phys.\ Rev.\  D {\bf 66} (2002) 044012
  [arXiv:hep-th/0204112].
\bibitem{Brigante008gz}
  M.~Brigante, H.~Liu, R.~C.~Myers, S.~Shenker and S.~Yaida,
  ``The Viscosity Bound and Causality Violation,''
  arXiv:0802.3318 [hep-th].
  A.~Buchel and R.~C.~Myers,
 ``Causality of Holographic Hydrodynamics,''  JHEP {\bf 0908} (2009) 016  [arXiv:0906.2922 [hep-th]].
  D.~M.~Hofman,
  ``Higher Derivative Gravity, Causality and Positivity of Energy in a UV complete QFT,''
    Nucl.\ Phys.\ B {\bf 823} (2009) 174  [arXiv:0907.1625 [hep-th]].

\bibitem{Kim:2007em}
  Y.~Kim, B.~-H.~Lee, S.~Nam, C.~Park and S.~-J.~Sin,
  ``Deconfinement phase transition in holographic QCD with matter,''
  Phys.\ Rev.\ D {\bf 76} (2007) 086003
  [arXiv:0706.2525 [hep-ph]].

\bibitem{Park:2009nb}
  C.~Park,
  ``The Dissociation of a heavy meson in the quark medium,''
  Phys.\ Rev.\ D {\bf 81} (2010) 045009
  [arXiv:0907.0064 [hep-ph]].
\bibitem{Fadafan:2012qy}
  K.~B.~Fadafan and E.~Azimfard,
  ``On meson melting in the quark medium,''
  Nucl.\ Phys.\ B {\bf 863} (2012) 347
  [arXiv:1203.3942 [hep-th]].
\bibitem{Fadafan:2008uv}
  K.~B.~Fadafan,
  ``Charge effect and finite 't Hooft coupling correction on drag force and Jet Quenching Parameter,''
  Eur.\ Phys.\ J.\ C {\bf 68} (2010) 505
  [arXiv:0809.1336 [hep-th]].
\bibitem{G1}
U. Gursoy, E. Kiritsis, L. Mazzanti and F. Nitti, "Deconfinement and
Gluon Plasma Dynamics in Improved Holographic QCD," Phys. Rev. Lett.
101 (2008) 181601 [27] S. S. Gubser, A. Nellore, S. S. Pufu and F.
D. Rocha, "Thermodynamics and bulk viscosity of approximate black
hole duals to finite temperature quantum chromodynamics," Phys. Rev.
Lett. 101 (2008) 131601
\bibitem{Lee:2013oya}
  B.~-H.~Lee, S.~Mamedov, S.~Nam and C.~Park,
  ``Holographic meson mass splitting in the Nuclear Matter,''
  JHEP {\bf 1308} (2013) 045  [arXiv:1305.7281 [hep-th]].



\end{thebibliography}
\end{document}